\documentstyle[12pt]{article}
\title{Evolutionary and asymptotic stability 
in symmetric multi-player games} 
\author{Maciej Bukowski \\
Institute of Economics \\
Warsaw School of Economics \\ 
Aleje Niepodleg\l o\'{s}ci 162 \\
02-554 Warsaw, Poland \\
e-mail: mbukows@sgh.waw.pl \\
and \\
Jacek Mi\c{e}kisz 
\\ Institute of Applied Mathematics and Mechanics 
\\ Warsaw University  \\ ul. Banacha 2  \\ 02-097
Warsaw, Poland 
\\ e-mail: miekisz@mimuw.edu.pl} 
\begin{document} 
\baselineskip=26pt
\maketitle 
 
\noindent {\bf Abstract.} We provide a classification 
of symmetric three-player games with two strategies and
investigate evolutionary and asymptotic stability 
(in the replicator dynamics) of their Nash equilibria. 
We discuss similarities and differences between two-player 
and multi-player games. In particular, we construct examples 
which exhibit a novel behavior not found in two-player games.  
\vspace{7mm}

\eject
 
\section{Introduction}
Equilibrium behavior of systems of many interacting entities can be
described in the framework of game-theoretic models. Although there are many
players in these models, their strategic interactions are usually decomposed
into a sum of two-player games \cite{kmr,blume1,ellis1,young,ellis2}.
Only recently, there have appeared some systematic studies 
of truly multi-player games \cite{kim,broom,koba}. 
 
We will provide here a classification of symmetric three-player games 
with two strategies and describe their symmetric Nash equilibria
and evolutionarily stable strategies. As in two-player games, 
for a certain range of payoff parameters, there exist multiple Nash
equilibria.
 
In the first class of our games, there are two pure Nash equilibria
and a mixed one. Such games are three-player analogs of two-player 
coordination games. In the second class, there are two mixed Nash
equilibria and a pure one. One may also have games with one pure 
and one mixed Nash equilibrium. We are faced therefore with a standard 
problem of equilibrium selection. We discuss in this context 
evolutionarily stable strategies. We develop a simple criterion 
to check whether a given Nash equilibrium is evolutionarily stable  
in a game with two strategies and apply it to three-player games. 
Then we investigate the asymptotic stability of Nash equilibria 
in the replicator dynamics. Here we also encounter some novel behavior. 
It concerns supersymmetric games, i.e., those where for any given profile 
of strategies, payoffs of all players are the same. Any symmetric $n$-player 
game with two strategies can be transformed by the standard payoff transformation 
into a supersymmetric game which has the same set of Nash equilibria 
and evolutionarily stable strategies \cite{hofpla,hofso,san}. In other words, 
any $n$-player game with two strategies is a so called potential game \cite{mon}. 
It is known that in two-player games, 
interior evolutionarily stable strategies are globally asymptotically
stable. Three-player games with two strategies belonging to the second
category of our classification have two evolutionarily stable strategies, 
a pure and an interior one. Due to a general theorem, 
they are both asymptotically stable and therefore the interior one 
is not globally stable. We also construct an example 
of a four-player supersymmetric game with two interior evolutionarily stable strategies, 
hence neither of them is globally stable. It is also known that 
in supersymmetric two-player games, a strategy is evolutionarily stable if and only if 
it is asymptotically stable in the replicator dynamics. 
We show that this is also true in supersymmetric n-player games. 
 
Finally, we discuss the concept of risk-dominance \cite{hs} 
and its relation to the size of the basin of attraction 
in the replicator dynamics of a Nash equilibrium. 
We show, that unlike in two-player games with two strategies, 
risk-dominant strategies may have smaller basins of attraction 
than dominated ones.

In Section 2, we define multi-player games. In Section 3, 
we provide a classification of symmetric three-player games with two strategies and
describe their symmetric Nash equilibria. In Section 4, we discuss evolutionarily stable 
strategies of our games. In Section 5, we study asymptotic stability of Nash equilibria 
in the replicator dynamics. In Section 6, relations between the risk-dominance
and the size of the basin of attraction are analyzed. Discussion follows in Section 7. 
 
\newtheorem{theo}{Theorem}
\newtheorem{defi}{Definition}
\newtheorem{hypo}{Hypothesis}
\newtheorem{example}{Example}
\newtheorem{corollary}{Corollary}
\newtheorem{lemma}{Lemma}
\newtheorem{prop}{Proposition}
 
\section{Multi-Player Games}
 
\begin{defi}
A {\bf game in the normal form} is a triple $G=\left(
I,S,\pi \right) $, where $I=\left\{
1,2,...,n\right\} $ is the set of players,
$S=\times_{i=1}^{n} S_{i}$, 
where $S_{i}=\left\{ 1,2,...,k_{i} \right\} $, 
is the finite set of strategies available for each player,
$\pi:S\rightarrow R^{n}$ is a payoff function assigning to every
profile of pure strategies, $s=(s_{1},s_{2},...,s_{n}) \in S $, a vector of payoffs, 
$\pi \left( s\right) =\left(
\pi _{1}\left( s\right) ,\pi _{2}\left( s\right) ,...\pi _{n}\left(
s\right) \right) $, where $\pi _{i}\left( s\right)$ is the payoff 
of the $i$-th player in the profile $s$.
\end{defi}
 
We will consider here only symmetric games. In such games, all players
assume the same role in the game and moreover the payoff 
of any player depends only on his strategy and numbers 
of players playing different types of strategies. More formally,
 
\begin{defi}
G is called symmetric, if for every permutation of the set of players,
$\sigma :I\rightarrow I$, we have $\pi _{i}\left(
s_{1},s_{2},...,s_{n}\right) =\pi _{\sigma ^{-1}\left( i\right) }\left(
s_{\sigma \left( 1\right) },s_{\sigma \left( 2\right) },...,s_{\sigma
\left(
n\right) }\right) $.
\end{defi}
 
A special subclass of symmetric games consist of supersymmetric games,
where all players get the same payoff dependent only on 
the profile of strategies.
 
\begin{defi}
$G=\left( I,S,\pi \right) $ is called supersymmetric if 
for every permutation of the set of players, 
$ \sigma:I\rightarrow I$, and every $i$, 
$\pi _{i}\left( s_{1},s_{2},...,s_{n}\right) =\pi
_{1}\left( s_{\sigma \left( 1\right) },s_{\sigma \left( 2\right)
},...,s_{\sigma \left( n\right) }\right) $.
\end{defi}
 
Payoffs in symmetric games are uniquely determined by the payoff
function of, say, the first player: $u_{1}: S \rightarrow R$.
We set $a_{s_{1}s_{2}...s_{n}}=\pi _{1}\left( s_{1},s_{2},...,s_{n}\right) $.
 
The payoff function of a two-player game with $k$ strategies 
can be then represented by a $k \times k$ matrix,
$A=(a_{ij})$, where $a_{ij}$ is the payoff of the first (row) 
player when he plays the strategy $i$, while the second (column)
player is playing the strategy $j$.
 
Payoffs of a three-player game with $k$ strategies can be represented 
by $k$ such matrices.
 
\begin{equation}
A=\left( {\small A}_{1},{\small A}_{2},...,{\small A}%
_{k}\right), 
\end{equation}
where 
$$\forall _{j\in \left\{ 1,2,...,k\right\} } \; \;
{\small A}_{j}{\small =}\left( 
\begin{array}{cccc}
a_{11j} & a_{12j} & ... & a_{1kj} \\ 
a_{21j} & a_{22j} & ... & a_{2kj}\\ 
... &  &  &  \\ 
a_{k1j} & a_{k2j} & ... & a_{kkj}
\end{array}
\right)$$ 
 
is a payoff matrix of the row player when the third (matrix) player
plays the strategy $j$. 

In particular, in the case of a three-player
game with two strategies
\begin{equation}
A=\left( \left( 
\begin{array}{cc}
a_{111} & a_{121} \\ 
a_{211} & a_{221}
\end{array}
\right) ,\left( 
\begin{array}{cc}
a_{112} & a_{122} \\ 
a_{212} & a_{222}
\end{array}
\right) \right). 
\end{equation}
 
Payoffs in a four-player game with two strategies are given by
 
\begin{equation}
{\small A}{\small =}\left( \left( \left( 
\begin{array}{cc}
a_{1111} & a_{1211} \\ 
a_{2111} & a_{2211}
\end{array}
\right) ,\left( 
\begin{array}{cc}
a_{1121} & a_{1221} \\ 
a_{2121} & a_{2221}
\end{array}
\right) \right) ,\left( \left( 
\begin{array}{cc}
a_{1112} & a_{1212} \\ 
a_{2112} & a_{2212}
\end{array}
\right) ,\left( 
\begin{array}{cc}
a_{1122} & a_{1222} \\ 
a_{2122} & a_{2222}
\end{array}
\right) \right) \right)   
\end{equation}
 
A {\bf mixed strategy} of the $i$-th player will be denoted by $x_{i}$
and the set of all mixed strategies by $\Delta_{i}$,   
$$\Delta_{i}=\{ x_{i} \in R^{k}:x_{i}=\{
x_{i1},x_{i2},...,x_{ik}\}, \; where \; x_{im}\geq 0 \; and \;
\sum_{m=1}^{k}x_{im}=1 \}.$$ 
A pure strategy $m$ is sometimes denoted by $e^{m}=(0,...,1,...,0)$.
The support of a mixed strategy $x_{i}$ is the set of pure strategies
played with nonzero probabilities,
$$supp(x_{i}) =\{ m\in S_{i}: x_{im}>0 \}$$
 
$\Theta=\times_{i=1}^{n}\Delta_{i}$ is the set of profiles of mixed
strategies. 
  
If $s=\{s_{1},...,s_{n}\}$ and $x\in \Theta$, 
then $x(s)=\prod_{i=1}^{n}x_{is_{i}}$
and we set the expected payoff for the $i$-th player
\begin{equation}
u_{i}(x) = \sum_{s \in S}x(s)\pi_{i}(s)
\end{equation}
 
\begin{defi}
$x \in \Theta$ is a {\bf Nash equilibrium} if
$\forall_{ i \in I, \; z_{i} \in \Delta_{i}}$
$$u_{i}(x_{i},x_{-i}) \geq u_{i}(z_{i},x_{-i}),$$
where $x_{-i} \in \times_{j \neq i} \Delta_{j}$ is a profile of
strategies of all players except the $i-$th one.
\end{defi} 
The set of Nash equilibria of any game with finite number of players 
and strategies is nonempty and is denoted by $\Theta^{NE}$.

Here we consider only symmetric Nash equilibria, that is elements of 
$$\Delta ^{NE}=\{ x\in \Delta:(x,x^{n-1}) \in \Theta^{NE}\},$$
where $\Delta$ is the simplex of mixed strategies common for all
players and $(x,x^{n-1})=(x,x,...,x).$
The payoff $u_{i}(x,x^{n-1})$ is the same for all $i$
and we denote it by $u(x,x^{n-1}).$
 
We have the following useful characterization of symmetric mixed Nash 
equlibria. 
 
\begin{theo}
In a symmetric game, if $x\in \Delta
^{NE}$, then $\forall _{m\in supp\left( x\right) }$ $u\left(
e^{m},x^{n-1}\right) =u\left( x,x^{n-1}\right) =const.$ If $x\in
int\,\Delta $,$\ $ then $x\in \Delta ^{NE}$ if and only if $\forall
_{m}$ $u\left( e^{m},x^{n-1}\right) =u\left( x,x^{n-1}\right) =const$.
\end{theo}
 
{\bf Proof}: If $x\in \Delta ^{NE}$, then $u\left(
e^{m},x^{n-1}\right) \leq u\left( x,x^{n-1}\right)$ for all $m$. 
If there existed $m\in supp\left( x\right) $ such that $u\left(
e^{m},x^{n-1}\right) <u\left( x,x^{n-1}\right) $, then 
$$u\left( x,x^{n-1}\right)= 
\sum_{m \in supp(x)}x_{i}u\left( e^{m},x^{n-1}\right)<
\sum_{m \in supp(x)}x_{i}u\left( x,x^{n-1}\right)=
u\left( x,x^{n-1}\right)$$
which is not possible. Similarly, if $x\in int\,\Delta $ and $%
\forall _{m}$ $u\left( e^{m},x^{n-1}\right) =u\left( x,x^{n-1}\right)
$, then
$\forall _{y\in \Delta }$
$$u\left( y,x^{n-1}\right)= 
\sum_{m \in supp(y)}y_{i}u\left( e^{m},x^{n-1}\right)=
\sum_{m \in supp(y)}y_{i}u\left( x,x^{n-1}\right)=
u\left( x,x^{n-1}\right)$$ 
which finishes the proof.
 
\section{Nash Equilibria in Three-Player Games with Two Strategies}
 
In this section, we completely characterize symmetric
Nash equilibria in symmetric three-player games 
with two strategies. Without loss of generality 
(adding  payoffs to columns of payoff matrices does not change 
the incentive functions and therefore neither Nash equilibria
nor their stability properties in the replicator dynamics)
we will consider games defined 
by the following generic payoff function:
$$\begin{array}{c}
A=\left( \left( 
\begin{array}{cc}
a & 0 \\ 
0 & b
\end{array}
\right) ,\left( 
\begin{array}{cc}
0 & 0 \\ 
b & c
\end{array}
\right) \right) 
\end{array}$$
We may also add $b$ to the fourth column to obtain a supersymmetric game
with the payoff function

$$\begin{array}{c}
A^{\prime}=\left( \left( 
\begin{array}{cc}
a & 0 \\ 
0 & b
\end{array}
\right) ,\left( 
\begin{array}{cc}
0 & b \\ 
b & c+b
\end{array}
\right) \right) 
\end{array}$$
Both $A$ and $A^{\prime}$ have the same family of Nash equilibria.

It follows that $\left( 1,0\right) \in \Delta ^{NE}$
if and only if $a\geq 0$ and $\left( 0,1\right) \in \Delta^{NE}$ 
if and only if $c\geq 0$. Now we will investigate symmetric 
interior (mixed) Nash equilibria. In games with two strategies 
we will identify $x \in \Delta$ with the probability 
of playing the first strategy.
Theorem 1 tells us that an interior point $(x,1-x)$ of the simplex
$\Delta$ is a symmetric Nash equilibrium if and only if 
\begin{equation}
ax^{2}=2bx\left( 1-x\right) +c\left( 1-x\right) ^{2}
\end{equation}
which is equivalent to the following equation: 
\begin{equation}
(a+2b-c)x^{2}-2(b-c)x-c=0
\end{equation}
If $b=\frac{\left(c-a\right) }{2}$, 
then the above equation is reduced to 
$$(a+c)x=c.$$
Hence if $a=-c \neq 0$, then there are no interior symmetric Nash equlibria,
if $a=b=c=0$, then every point is a Nash equilibrium and finally if
$a\neq -c$ and $ac>0$, then $x=\frac{c}{a+c}$
is an interior Nash equilibrium. Now we will consider the case 
when $w\left( x\right) $ is a quadratic function. Let us notice that 
$$\begin{array}{cl}
i) & w\left( 0\right) =-c \; and \; w\left( 1\right) = a \\ 
ii) & w^{\prime }\left( x_{e}\right) =0 \Leftrightarrow x_{e}=
\frac{b-c}{a+2b-c} \\ 
iii) & w\left( x_{e}\right) =-\frac{b^{2}+ac}{a+2b-c}
\end{array}$$
where $w^{\prime}(x)$ is the derivative of $w(x)$.
From $\left( i\right)$ we have that if $ac>0$, then 
$w\left( x\right)$ has a unique root in the interval 
$\left( 0,1\right)$ hence there exists a unique symmetric interior Nash
equilibrium. However, if $ac<0$, then an interior symmetric Nash equilibrium
exists if and only if the following conditions are satisfied:
$$\begin{array}{cl}
i) & 0<x_{e}<1, \\ 
ii) & w\left( 0\right) w\left( x_{e}\right)
=c\frac{b^{2}+ac}{a+2b-c}\leq 0.
\end{array}$$
From $\left( i\right)$ we obtain that either
$b>\max \left(c,-a\right) =-\min \left(
-c,a\right) $
or $b<\min \left(c,-a\right) =-\max \left(
-c,a\right) ,$ which together with $\left( ii\right) $
gives us that if $ac<0$, then a symmetric interior Nash equilibrium exists
if and only if one of two following conditions is satisfied: 
$$\begin{array}{cl}
a) & c<0 \; and \; b\geq \sqrt{|ac|} \\ 
b) & c>0 \; and \; b\leq -\sqrt{|ac|}
\end{array}$$
Moreover, if $\left| b\right| =\sqrt{|ac|}$, 
then there exists a unique
symmetric interior Nash equilibrium and if 
$\left| b\right| >\sqrt{|ac|}$,
then there are two such Nash equilibria. 
Now we have to deal with the case
$ac=0.$ If $a=0$, then $w\left( x\right) $
has the following form: 
$$\begin{array}{c}
w\left( x\right) =x^{2}\left( 2b-c\right) -2x\left(
b-c\right) -c \\ 
w^{\prime }\left( x\right) =2x\left( 2b-c\right) -2\left(
b-c\right) 
\end{array}$$
which means that $w\left( 1\right) =0$ and $w^{\prime }\left( 1\right)
=2b$ hence $w\left( x\right) $ has a root in $\left( 0,1\right)$ 
if and only if $bc<0.$ If $c=0$, then
$w\left( x\right) $ is reduced to: 
\begin{eqnarray}
w\left( x\right)  &=&x^{2}\left(a+ 2b\right) -2bx, \\
w^{\prime }\left( x\right)  &=&2x\left(a+ 2b\right) -2b
\end{eqnarray}
and it follows that $w\left( 0\right) =0$ and $w^{\prime }\left(
0\right) =-2b$
hence $w\left( x\right)$ has a root in $\left( 0,1\right)$ if and only 
if $ab>0.$ 
We can sum up the above analysis in the following classification
of three-player games with two strategies. \medskip 
 
\textbf{Category I:} Games with three symmetric (two pure and one
mixed) Nash equilibria.
 
$\Delta ^{NE}=\{ \left( 1,0\right) ,\left( 0,1\right) ,\left(
x,1-x\right) \} $ if $( a>0$ and $c>0) $ 
or $( a=0, b<0,$ and $c>0) $ or $(a>0, b>0,$ and $c=0) $. 
In the first case, if $b\neq
\frac{\left(c-a\right) }{2}$, then $x=\frac{\left(
b-c\right) +\sqrt{b^{2}+ac}}{\left(
a+2b-c\right) }$ and if $b=\frac{\left(
c-a\right) }{2}$, then $x=\frac{c}{a+c}$ . In the second case, $x
=\frac{-c}{2b-c}$. In the third case, $x=\frac{2b}{2b+a}$.\medskip 
 
\textbf{Category II:} Games with three symmetric (one pure and two
mixed) Nash equilibria.
 
\begin{enumerate}
\item  $\Delta ^{NE}=\{ (1,0),
(x_{1},1-x_{1}) ,( x_{2},1-
x_{2}) \} $, where $x_{1}=\frac{\left(
b-c\right) -\sqrt{b^{2}+ac}}{\left(
a+2b-c\right) }$ and $x_{2}=\frac{\left(
b-c\right) +\sqrt{b^{2}+ac}}{\left(
a+2b-c\right) }$, if $a>0$, $
c<0$ and  $b>\sqrt{|ac|}$.
 
\item  $\Delta^{NE}=\{ (0,1), (x_{1},1-x_{1}),
(x_{2},1-x_{2}) \} $, where $x_{1}$ and $x_{2}$ are as before, 
if $a<0$,\ $ c>0$ and  $b<-\sqrt{|ac|}$. \medskip 
\end{enumerate}
 
\textbf{Category III:} Games with one pure and one mixed Nash 
equilibrium. \medskip 
 
\begin{enumerate}
\item  $\Delta ^{NE}=\{ \left( 1,0\right)
,\left( x,1-x\right) \} $ if 
$( a>0, c<0,$ and $b=\sqrt{|ac|}) $ 
or $( a=0$, $b>0,$ and $c<0). $
In the first case, $x=\frac{\sqrt{|c|}}{\sqrt{a}+%
\sqrt{|c|}}$. In the second case,
$x=\frac{-c}{2b-c}$.
 
\item $\Delta ^{NE}=\{ \left( 0,1\right)
,\left( x,1-x\right) \} $ if 
$( a<0, c>0$  and $b=-\sqrt{|ac|}) 
$ or $(a<0, b<0,$ and $c=0) $.
In the first case, $x=\frac{\sqrt{c }
}{\sqrt{|a| }+\sqrt{c}}$. In the
second case,
$x=\frac{2b}{2b+a}$.\medskip 
\end{enumerate}
 
\textbf{Category IV:} Games with one mixed Nash equilibrium.
 
$\Delta ^{NE}=\{\left( x,1-x\right) \} $ 
if $a<0$ and $c<0$.
If $b=\frac{\left(c-a\right) }{2}$, then
$x=\frac{c}{a+c}$ and if $b\neq \frac{
\left(c-a\right) }{2}$, then $x=\frac{\left(
b-c\right) -\sqrt{b^{2}+ac}}{\left(
a+2b-c\right) }$.\medskip 
 
\textbf{Category V:} Games with two pure Nash equilibria.
 
$\Delta^{NE}=\{ \left( 1,0\right) ,\left( 0,1\right) \} $ 
if $( a=0, b\geq 0,$ and $c>0) $ or $(a>0, b\geq
0$, and $c=0) $ or $( a=0, b\neq 0)$, and $c=0$. \medskip
 
\textbf{Category VI:} Games with one pure Nash equilibrium.
 
\begin{enumerate}
 
\item  $\Delta^{NE}=\{ \left( 1,0\right)
\} $ if $( a>0, c<0,$ and $b<\sqrt{|ac|}) $ or
$( a=0, b\leq 0,$ and $c<0) $.
 
\item  $\Delta ^{NE}=\{ \left( 0,1\right)
\} $ if $( a<0, c>0,$ and $b>-\sqrt{|ac|})$ 
or $(a<0, b\leq 0,$ and $c=0) $.\medskip 
\end{enumerate}
 
\textbf{Category VII:} Games with $\Delta ^{NE}=\Delta $ for
$a=0$,\ $b=0,$ and $c=0 $.\medskip
 
Let us notice that Categories III, IV, and VII contain nongeneric games.

In five of the above categories there are multiple Nash equilibria. 
We are therefore faced with the problem of equilibrium selection.
 
\section{Evolutionarily Stable Strategies}
 
We introduce the following definition of an evolutionarily stable strategy 
for n-player games \cite{broom,koba}.
\begin{defi}
$x\in \Delta $ is an {\bf evolutionarily stable strategy} (ESS)
if there exists $0 < \epsilon^{'} <1$ such that for all $0< \epsilon <\epsilon^{'}$
and for every $y \in \Delta$
$$u\left( x,\left( \epsilon y+\left(
1-\epsilon \right) x\right) ^{n-1}\right) > u\left( y,\left(
\epsilon y+\left( 1-\epsilon \right) x\right) ^{n-1}\right).$$ 
\end{defi}
 
\begin{defi}
\label{def_lok_nadrz} $x\in \Delta $ is {\bf locally superior} if
there exists a neighborhood $U$ of $x$ such that for all 
$y\in U\cap \Delta $ we have 
\begin{equation}
u\left( x,y^{n-1}\right) \geq u\left( y,y^{n-1}\right)
\label{lokdomess}
\end{equation}
and the equality holds only if $y=x$.
\end{defi}
 
The following theorem is a straightforward generalization of
the corresponding theorem for two-player games \cite{hofsi}. 
 
\begin{theo}
\label{tw_lok_nadrz} $x \in \Delta^{ESS}$ if and only if $x$ is locally superior.
\end{theo}
 
The above theorem can be used to obtain a useful criterion 
for evolutionary stability in the case 
of symmetric $n$-player games with two strategies \cite{broom}. 
Let $x,y \in \Delta .$ Define an incentive function: 
$$w\left( y\right) =u\left(
e^{1},y^{n-1}\right)
-u\left( e^{2},y^{n-1}\right).$$
\begin{theo}
In symmetric n-player games with two strategies,
$x\in \Delta ^{ESS}$ if and only if 
for $y\in \Delta $ close to $x$ both of the
following implications hold:
\begin{eqnarray}
If \; \; y &<&x, then \; w\left( y\right) >0, \\
If \; \; y &>&x, then \; w\left( y\right) <0.
\end{eqnarray}
\end{theo}
 
\textbf{Proof: } Consider the function 
$$w_{x}\left( y \right) =u\left( x-y,y^{n-1}\right).$$
Due to Theorem \ref{tw_lok_nadrz},  $x \in \Delta ^{ESS}$ 
if and only if for any $y \in \Delta $
close to $x$ and $y \neq x$
we have $w_{x}\left( y\right) >0$. It follows that 
$$w_{x}\left( y\right) =\left( x-y\right)
\left(u\left( e^{1},y^{n-1}\right) -u\left( e^{2},y^{n-1}\right) \right)$$
which finishes the proof.

It follows that $x\in \Delta $ is evolutionarily stable if and only if
$w\left( y \right)$ is a decreasing function in the neighborhood of $x$. 
 
For $n \leq 3$ we have the following corollary. 

\begin{corollary}
\label{wniosloknad} $x \in \Delta ^{ESS}\cap int$ $\Delta $ 
if and only if $w\left( x\right) =0\ $ and $w^{\prime }\left( x\right) <0 $.
\end{corollary}
 
The following theorems concern the number and supports 
of evolutionarily stable strategies.
  
\begin{theo}
\label{tw_Bish_Can} In symmetric two-player games, if $x,y\in
\Delta ^{ESS}$ and $x \neq y$, then $supp\left( x\right) 
\not\subseteq supp\left( y\right) $ \cite{bish1}.
\end{theo}
 
In particular, if $x\in \Delta ^{ESS}\cap int$ $\Delta $, then $x$ 
is the unique evolutionarily stable strategy. 
The analogous theorem is not true in three-player games; 
games in Category I of our classification have one pure and one interior 
evolutionarily stable strategy. For three-player games we have
the following weaker theorem.
 
\begin{theo}
\label{tw_Broom} 
In symmetric three-player games, if $x,y\in \Delta ^{ESS}$ and
$x \neq y$, then $supp\left( x\right)
\neq supp\left( y\right) $ \cite{broom}.
\end{theo}
 
It was already indicated in \cite{broom} that the above theorem cannot be
generalized to games with $n\geq 4$ players. We illustrate it with the following example.
\vspace{3mm}
 
\noindent {\bf Example 1} A 4-player supersymmetric game with two interior 
evolutionarily stable strategies
\vspace{3mm}
 
\noindent Payoffs are given by 
$$\begin{array}{c}
A=\left( \left( \left( 
\begin{array}{cc}
a_{1} & 0 \\ 
0 & a_{2}
\end{array}
\right) ,\left( 
\begin{array}{cc}
0 & a_{3} \\ 
a_{2} & 0
\end{array}
\right) \right) ,\left( \left( 
\begin{array}{cc}
0 & a_{3} \\ 
a_{2} & 0
\end{array}
\right) ,\left( 
\begin{array}{cc}
a_{3} & 0 \\ 
0 & a_{4}
\end{array}
\right) \right) \right)
\end{array}$$
If $x \in \Delta ^{NE}\cap int$ $%
\Delta $, then $x$ is a root of the equation $w\left( x\right) =0$, where 
$$w\left( x\right) =a_{1}x^{3}+3a_{3}x\left( 1-x\right)^{2}-3a_{2}x^{2}\left(
1-x\right) -a_{4}\left( 1-x\right) ^{3}.$$
Put $a_{1}=a_{4}=-\frac{3}{32}$ and 
$a_{2}=a_{3}=-\frac{13}{96}$. Then: 
\begin{equation}
w(x)=-\left( x-\frac{1}{4}\right) \left(
x-\frac{1}{2}\right)
\left( x-\frac{3}{4}\right)
\end{equation}
It follows from Theorem 4 that both $\left(
x_{1},1-x_{1}\right) =\left( \frac{1}{4},\frac{3}{4}\right) $ and
$\left(x_{2},1-x_{2}\right) =\left( \frac{3}{4},\frac{1}{4}\right) $ 
are interior evolutionarily stable strategies.
 
\subsection{Evolutionarily stable strategies in three-player games with
two strategies}
 
In this section, we completely characterize 
evolutionarily stable strategies in symmetric three-player games 
with two strategies. Again, without loss of generality we will consider
games defined by the following generic payoff function: 
\begin{equation}
\begin{array}{c}
A=\left( \left( 
\begin{array}{cc}
a & 0 \\ 
0 & b
\end{array}
\right) ,\left( 
\begin{array}{cc}
0 & 0 \\ 
b & c
\end{array}
\right) \right)
\end{array}
\label{post_norm}
\end{equation}
Due to Theorem 3, $x \in \Delta$ is ESS 
if and only if for all
$y=\left( y,1-y\right) \in \Delta $ close to $x$ 
the following conditions hold: 
\begin{eqnarray}
y &<&x \Leftrightarrow ay^{2}-2by\left( 1-y\right)
 -c\left( 1-y\right)^{2} >0, \\
y &>&x \Leftrightarrow ay^{2}-2by\left( 1-y\right)
 -c\left( 1-y\right)^{2} <0.
\end{eqnarray}
 
We will deal with the following polynomial: 
\begin{equation}
w\left( y\right) =y^{2}\left( a+2b-c \right)
-2y\left(b-c\right) -c  \label{troj_ess_wiel}.
\end{equation}
$x \in \Delta $ is ESS if and only if
for all $y$ close to $x$ and $y<x$ we have $w\left(
y\right) >0$ and for $y>x$, $w\left(
y\right) <0$. Checking the derivative of $w(y)$ 
we arrive at the following classification.
\medskip
 
\textbf{Category I:} $\Delta ^{ESS}=\{ \left( 1,0\right) ,\left(
0,1\right) \} $ and $\Delta ^{NE}\backslash \Delta ^{ESS}=\left\{
\left( x,1-x\right) \right\} $.\medskip
 
\textbf{Category II:} 
 
\begin{enumerate}
\item  $\Delta ^{ESS}=\{ \left( 1,0\right) ,\left( 
x_{1},1-x_{1}\right) \} $ and $\Delta
^{NE}\backslash \Delta ^{ESS}=\left\{ \left( x_{2},1-x_{2} 
\right) \right\} $.
 
\item  $\Delta ^{ESS}=\{ \left( 0,1\right) ,\left( 
x_{1},1-x_{1}\right) \} $ and $\Delta
^{NE}\backslash \Delta ^{ESS}=\left\{ \left(
x_{2},1-x_{2} \right) \right\} $.\medskip
\end{enumerate}
 
\textbf{Category III:}
\begin{enumerate}
\item  $\Delta ^{ESS}=\{ \left( 1,0\right) \} $
and $\Delta ^{NE}\backslash \Delta ^{ESS}=\left\{ \left(
x,1-x \right) \right\} $.
 
\item $\Delta ^{ESS}=\{ \left( 0,1\right) \} $
and $\Delta ^{NE}\backslash \Delta ^{ESS}=\left\{ \left(
x,1-x \right) \right\} $.\medskip
\end{enumerate}
 
\textbf{Category IV:} $\Delta
^{ESS}=\{ \left( x,1-x \right) \} $ and $%
\Delta ^{NE}\backslash \Delta ^{ESS}= \emptyset$.\medskip
 
\textbf{Category V:} If $a=0$, $b \geq 0$ and 
$c>0$, then $\Delta^{ESS}=\{ \left( 0,1\right) \} $ and 
$\Delta ^{NE}\backslash \Delta ^{ESS}=\{ \left( 1,0\right) \} $.
If $a>0$, $b\geq 0$, and $c=0$, then $\Delta ^{ESS}=\{
\left( 1,0\right) \} $ and $\Delta ^{NE}\backslash \Delta
^{ESS}=\{ \left( 0,1\right) \} $. If  $a=0$,  $b<0$,
and $c=0$, then $\Delta ^{ESS}=\{ \left( 0,1\right) \} $ and $
\Delta ^{NE}\backslash \Delta ^{ESS}=\left\{ \left( 1,0\right) \right\}$.
If $a=0$, $b>0$, and $c=0$, then $\Delta
^{ESS}=\{ \left( 1,0\right) \} $ and $\Delta ^{NE}\backslash
\Delta ^{ESS}=\left\{ \left( 0,1\right) \right\} $.\medskip
 
\textbf{Category VI:} 
\begin{enumerate}
\item  $\Delta ^{ESS}=\{ \left( 1,0\right) \} $
and $\Delta ^{NE}\backslash \Delta ^{ESS}= \emptyset$.
 
\item  $\Delta ^{ESS}=\{ \left( 0,1\right) \} $
and $\Delta ^{NE}\backslash \Delta ^{ESS}= \emptyset$.\medskip
\end{enumerate}
 
\textbf{Category VII:}  $\Delta ^{NE}\backslash \Delta
^{ESS}=\Delta $.

\section{Asymptotic Stability in Replicator Dynamics}
 
The continuous replicator dynamics can be used to model the evolution 
of an infinite population. The state of the population is characterized
by the vector $x(t)=(x_{1},...,x_{k})$, where for every $m$, $x_{m}$ 
is the fraction of individuals playing the strategy
$m$. Players are repeatedly matched with other $n-1$ players to play
an $n$-player stage game. Formally, $x(t)\in \Delta$ 
and we may naturally interpret $x$ as a mixed strategy 
of a single player. A system of $k$ differential equations
 
\begin{equation}
\dot{x}_{m}=x_{m}\left( u_{m}\left( x\right) -\sum_{l=1}^{k} x_{l}u_{l}\left(
x\right) \right), m=1,...,k  \label{replikator1}
\end{equation}
was proposed by Taylor and Jonker \cite{jonktay} as a dynamical model
for two-player games. For $n$-player games,  replicator dynamics 
can be written in the following way: 
 
\begin{equation}
\dot{x}_{m}=x_{m}\left[ u\left( e^{m},x^{n-1}\right) -u\left(
x,x^{n-1}\right) \right].  \label{replikator2}
\end{equation}
Properties of replicator dynamics in two-player games have been studied
thoroughly. Some of them are also true for multi-player games. 
We will discuss these common features below. We will also present some
novel behavior in three and four-player games.
 
\subsection{Basic theorems}
 
We present here without proofs some fundamental theorems which were 
originally formulated for general $n$-player games and those whose 
proofs can be easily generalized from two-player to multi-player games.
We will also show that in symmetric three-player games 
with two strategies, evolutionarily strategies are precisely those which
are asymptotically stable in replicator dynamics.
 
Let us denote the set of stationary points 
of the replicator dynamics (18) by 
$$\Delta ^{\ast }=\left\{ x\in \Delta :u\left( e^{m},x^{n-1}\right)
=u\left(
x,x^{n-1}\right) \forall _{m\in supp\left( x\right) }\right\},$$
interior stationary points by 
$\Delta ^{\ast \ast }=\Delta ^{\ast }\cap int\,\Delta ,$ stationary 
points stable in the Lyapunov sense by $\Delta ^{LSE}$, and
asymptotically stable points by
$\Delta ^{ASE}$. Let $x\left(t;x_{0}\right) $ be the solution of (18)
with the initial condition $x_{0}$
at time $t=0$, and $V$ and $U$ be open sets, then we have  
$$\begin{array}{l}
\Delta ^{LSE}=\left\{ x\in \Delta ^{\ast }:\forall _{V \ni x}
\exists _{U \ni x}\forall _{x_{0}\in U} x\left( 0;x_{0}\right)
\in U\Rightarrow \forall _{t\geq 0} x\left( t;x_{0}\right) \in
V\right\} \\ 
\Delta ^{ASE}=\left\{ x\in \Delta ^{LSE}:\exists _{V \ni x}
\forall _{U \ni x}\forall _{x_{0}\in V} \exists _{T\geq
0}\forall _{t\geq T} x\left( t;x_{0}\right) \in U\right\}
\end{array}$$
 
\begin{theo}
\label{tw_rep1}$\Delta ^{NE}\cup \left\{ e^{1},e^{2},...,e^{k}\right\}
\subset \Delta ^{\ast }$. Moreover if $x\in \Delta ^{\ast }$ is
Lyapunov stable, then $x\in \Delta ^{NE}$ \cite{weib}.
\end{theo}
 
\begin{theo}
\label{tw_rep2} If $x\left( 0\right) \in int\,\Delta $ and 
$\lim_{t \rightarrow \infty}x(t)=\hat{x}$, 
then $\hat{x} \in \Delta ^{NE}$ \cite{weib}.
\end{theo}

The following theorem was first proved for $n=2$ in \cite{hofsi}
and for general $n$ in \cite{palm}.
 
\begin{theo}
\label{tw_rep3} If $x\in \Delta ^{ESS}$, then $x$ is asymptotically
stable in the replicator dynamics, i.e., $x\in \Delta^{ASE}.$ 
\end{theo}

\begin{theo}
\label{tw_Nat_Sel} In supersymmetric $n$-player games, the average payoff
is increasing along any solution of the replicator dynamics equations.
It means that $\dot{u}\left( x,x^{n-1}\right) \geq 0$, 
and the equality holds only for $x\in $ $\Delta ^{\ast }$
(Prop. 3.14 in \cite{weib} and Th. 7.8.1 in \cite{hofb}).
\end{theo}

\subsection{New results}

Hofbauer and Sigmund provided an example of a symmetric two-player game
with three strategies and an asymptotically stable stationary point 
which is not evolutionarily stable \cite{hofb}.
We will show that in $n$-player supersymmetric games,
there is a one-to-one correspondence  
between asymptotically stable points 
and evolutionarily stable strategies. 
\begin{theo}
In supersymmetric n-player games, $x\in \Delta^{ESS}$ 
if and only if  $x\in \Delta ^{ASE}$ .
\end{theo}

We will adapt the proof for two-player supersymmetric games given in \cite{hofb}
(Th. 7.8.1), see also p.941 in \cite{broom}.

{\bf Proof:} Let $x$ be an asymptotically stable stationary point
of the replicator dynamics. By Theorem 9, $u(x,x^{n-1})=u(x^{n})$ attains a local
maximum at $x$. We will show that $x$ is locally superior and therefore
is evolutionarily stable. Let $x + \epsilon^{*} = 
(x_{1}+\epsilon_{1},...,x_{k}+\epsilon_{k})$,
$\sum_{i=1}^{k}\epsilon_{i}=0$, be a mixed strategy in the neighborhood of $x$.
If for every $i$, $|\epsilon_{i}| < \epsilon$ and $\epsilon$ is small enough, then
\begin{equation}
u((x+\epsilon^{*})^{n})<u(x^{n}).
\end{equation}
We expand the left-hand side of (19) and obtain
\begin{equation}
u((x+\epsilon^{*})^{n})=u(x+\epsilon^{*},(x+\epsilon^{*})^{n-1})=
u(x^{n})+\sum_{k=1}^{n-1}m_{k}u(x^{n-k}(\epsilon^{*})^{k})
+u((\epsilon^{*})^{n}),
\end{equation}
where $m_{k}=\left( \begin{array}{c} n-1 \\ k \end{array} \right)+
\left( \begin{array}{c} n-1 \\ k-1 \end{array} \right) = 
\left( \begin{array}{c} n \\ k \end{array} \right)$
is a binomial coefficient.
It follows, for sufficiently small $\epsilon$, 
that there exists $0<l\leq n$ such that
$u(x^{n-k}(\epsilon^{*})^{k})=0$ for $1\leq k<l$ and
$u(x^{n-l}(\epsilon^{*})^{l})<0$. 
We also have
\begin{equation}
u(x,(x+\epsilon^{*})^{n-1})=u(x^{n})+
\sum_{k=1}^{n-1}m_{k}^{'}u(x^{n-k}(\epsilon^{*})^{k}),
\end{equation}
where $m_{k}^{'}=\left( \begin{array}{c} n-1 \\ k \end{array} \right)$.
From (20-21), the remark after (20), and the fact that $m_{k} > m_{k}^{'}$ 
it follows, for sufficiently small $\epsilon$, 
that $$u((x+\epsilon^{*})^{n})<u(x,(x+\epsilon^{*})^{n-1})$$
so $x$ is locally superior and therefore evolutionarily stable.
The other direction is provided by Theorem 8.
\vspace{3mm}
 
Since any symmetric $n$-player game with two strategies
is equivalent to a supersymmetric game \cite{hofpla,hofso,san},
we have the following corollary.

\begin{corollary}
In symmetric $n$-player games with two strategies, $x \in \Delta^{ESS}$
if and only if $x \in \Delta^{ASE}.$ 
\end{corollary}

In particular, in Category I and II, there are two asymptotically 
stable Nash equilibria which are also evolutionarily stable;
in all other (except the last one) categories there is a unique
evolutionarily stable strategy which is globally asymptotically stable. 
Let us mention that in Category III there is a mixed Nash equilibrium 
which is stable from the right side and unstable from the left side 
and in Category V there is a pure unstable Nash equilibrium.
 
For two-player games we have
 
\begin{theo} 
If $x$ and $y$\ are two different interior stationary points of two-player
replicator dynamics, i.e., $x,y\in \Delta ^{\ast \ast }$, then for all 
$\alpha ,\beta \in R$ such that $\alpha x+\beta y\in \Delta $ 
we have $\alpha x+\beta y\in \Delta^{\ast \ast }$ \cite{zee}.
\end{theo}
 
This is not generally true in multi-player games; three-player games
in Category II of our classification have two isolated interior Nash equilibria.
 
In two-player games, every interior evolutionarily
stable strategy is globally asymptotically stable. 
\begin{theo} 
In two-player games, if
$\hat{x}\in int$ $\Delta \cap \Delta ^{ESS}$ and if 
$x\left(t,x_{0}\right) $ is a solution of a replicator dynamics passing by
$x_{0}\in int \Delta $, 
then $\lim_{t\rightarrow \infty}x\left( t,x_{0}\right) =\hat{x}$ \cite{hofb}.
\end{theo}
 
Example 1 shows that in supersymmetric 4-player games with 
two strategies there can be two interior evolutionarily stable strategies
and hence neither of them is globally asymptotically stable.
Let us also recall that games in Category II 
have two evolutionarily stable strategies,
one pure and one interior, so again the interior one
is not globally asymptotically stable. 

\section{Risk-Dominance and Asymptotic Stability} 
 
In the first two categories of our classification, there are games 
with two evolutionarily stable Nash equilibria. The problem of equilibrium selection 
is therefore not resolved. Here we will discuss the concept of risk-dominance
\cite{hs} and its relation to the size of the basin 
of attraction of a Nash equilibrium in the replicator dynamics.

The notion of risk-dominance was introduced and thoroughly studied 
by Hars\'{a}nyi and Selten \cite{hs}. The general theory is based 
on the so called tracing procedure. In symmetric 
two-player games with two strategies
and two symmetric pure Nash equilibria,
one can show that a strategy risk-dominates the other one 
if it has a higher expected payoff 
against a player playing both strategies with the probability $1/2$. 
It follows that this is equivalent to the first strategy
having a bigger basin of attraction.

The procedure of Hars\'{a}nyi and Selten was applied by Kim \cite{kim} 
to $n$-player games with two strategies 
and two pure symmetric Nash equilibria (three-player games 
in Category I of our classification; we assume that $a>0$ and $c>0$). 
He showed that the first strategy risk dominates the second one 
if and only if
\begin{equation}
a^{3}>2abc+c^{3}.    
\end{equation}
Now, the first strategy has a bigger basin of attraction than the second one
if $x <1/2$ ($x$ is the unique interior Nash equilibrium). 
This holds if the expected payoff 
of the first strategy is bigger than that of the second one
against two players playing both strategies with the probability $1/2$.  
We obtain that the first strategy has a bigger basin of attraction 
than the second one if and only if
\begin{equation}
a>2b+c.
\end{equation}
We have the following simple proposition.
\begin{prop}
Let $b>0$. If the first strategy has a bigger basin of attraction
than the second one, then it risk-dominates the second one.  
\end{prop}
{\bf Proof:} Assume that 
$$a-c>2b.$$
Then
$$(a-c)(a^{2}+ac+c^{2})>2b(a^{2}+ac+c^{2}),$$
hence
$$a^{3}-c^{3}>2abc.$$
If $b<0$, then of course an analogous proposition holds for the second strategy.

It is easy to see that if $b=0$ or $b=(c-a)/2$ 
(in the second case the incentive function $w(x)$ in (6) is linear), 
then conditions (22) and (23) are equivalent and risk-dominant strategies 
are those with bigger basins of attraction. 

However, as the following example shows, the converse proposition
does not hold.

\noindent {\bf Example 2} \hspace{5mm} If $a=4, b=1$, and $c=3$, 
then the first strategy is risk-dominant but it has 
a smaller basin of attraction than the second one.

The relationship between different approaches to equilibrium selection 
was recently discussed by Kim \cite{kim}. He provided an interpretation 
of the fact that in two-player games with two strategies, the risk-dominant 
strategy is selected by all criteria. He showed that that the equivalence
of those criteria breaks down for games with more than two players.

\section{Summary}
 
We provided a classification of symmetric three-player games
with two strategies and studied evolutionary and asymptotic
stability (in the replicator dynamics) of their Nash equilibria.
There exist two classes of games with two evolutionarily stable strategies.
In the first one, there are two pure evolutionarily stable strategies;
in the second one, there is one pure and one mixed evolutionarily stable strategy.
In subsequent papers we will investigate the stochastic stability 
of these Nash equilibria in stochastic replicator dynamics, 
Kandori-Mailath-Rob and Young models {and in spatial games with local interactions.
The problem of the selection of a mixed Nash equilibrium is especially interesting.  
\vspace{4mm}

\noindent {\bf Acknowledgments} JM would like to thank 
the Polish Committee for Scientific Research for a financial support
under the grant KBN 5 P03A 025 20.

\end{document}